\documentclass[twocolumn,prb,superscriptaddress,preprintnumbers,amsmath,amssymb,showpacs,floatfix]{revtex4}

\usepackage{graphicx}
\usepackage{dcolumn}
\usepackage{bm}
\usepackage{color}
\usepackage{multirow}
\usepackage{longtable}

\newcommand{\tg}{$t_{2g}$}
\newcommand{\eg}{$e_g$}
\newcommand{\led}{$L_{2,3}$}

\newcommand{\Mn}{MnWO$_4$}
\newcommand{\Co}{CoWO$_4$}

\newcommand{\Coz}{Co$^{2+}$}
\newcommand{\Mnz}{Mn$^{2+}$}

\begin{document}

\title{Local symmetry and magnetic anisotropy in multiferroic MnWO$_4$
and antiferromagnetic CoWO$_4$ studied by soft x-ray absorption
spectroscopy}

\author{N.~Hollmann}
 \affiliation{II. Physikalisches Institut, Universit\"{a}t zu K\"{o}ln,
 Z\"{u}lpicher Str. 77, 50937 K\"{o}ln, Germany}
\author{Z.~Hu}
 \affiliation{II. Physikalisches Institut, Universit\"{a}t zu K\"{o}ln,
 Z\"{u}lpicher Str. 77, 50937 K\"{o}ln, Germany}
 \affiliation{Max Planck Institute for Chemical Physics of Solids,
 N\"othnitzerstr. 40, 01187 Dresden, Germany}
\author{T.~Willers}
 \affiliation{II. Physikalisches Institut, Universit\"{a}t zu K\"{o}ln,
 Z\"{u}lpicher Str. 77, 50937 K\"{o}ln, Germany}
\author{L.~Bohat\'y}
 \affiliation{Institut f\"{u}r Kristallographie, Universit\"{a}t zu K\"{o}ln,
 Z\"{u}lpicher Str. 49b, 50674 K\"{o}ln, Germany}
\author{P.~Becker}
 \affiliation{Institut f\"{u}r Kristallographie, Universit\"{a}t zu K\"{o}ln,
 Z\"{u}lpicher Str. 49b, 50674 K\"{o}ln, Germany}
\author{A.~Tanaka}
 \affiliation{Department of Quantum Matter, ADSM, Hiroshima University,
 Higashi-Hiroshima 739-8530, Japan}
\author{H.~H.~Hsieh}
 \affiliation{Chung Cheng Institute of Technology,
 National Defense University, Taoyuan 335, Taiwan}
\author{H.-J.~Lin}
 \affiliation{National Synchrotron Radiation Research Center,
 101 Hsin-Ann Road, Hsinchu 30077, Taiwan}
\author{C.~T.~Chen}
 \affiliation{National Synchrotron Radiation Research Center,
 101 Hsin-Ann Road, Hsinchu 30077, Taiwan}
\author{L.~H.~Tjeng}
 \affiliation{II. Physikalisches Institut, Universit\"{a}t zu K\"{o}ln,
 Z\"{u}lpicher Str. 77, 50937 K\"{o}ln, Germany}
 \affiliation{Max Planck Institute for Chemical Physics of Solids,
 N\"othnitzerstr. 40, 01187 Dresden, Germany}

\date{\today}

\begin{abstract}
Soft x-ray absorption experiments on the transition metal $L_{2,3}$
edge of multiferroic MnWO$_4$ and antiferromagnetic CoWO$_4$ are
presented. The observed linear polarization dependence, analyzed
by full-multiplet calculations, is used to determine the ground
state wave function of the magnetic Mn$^{2+}$ and Co$^{2+}$ ions. The
impact of the local structure and the spin-orbit coupling on the
orbital moment is discussed in terms of the single-ion anisotropy.
It is shown that the orbital moment in CoWO$_4$ is responsible for the
collinear antiferromagnetism, while the small size of spin-orbit
coupling effects make spiral magnetic order in MnWO$_4$ possible,
enabling the material to be multiferroic.
\end{abstract}


\maketitle

The combination of magnetic order and ferroelectricity in
multiferroic compounds is a topic that has gained high interest in
modern research.\cite{fiebig05} Magnetic transitions directly
linked to a spontaneous electrical polarization have been
investigated in detail for materials like \emph{R}MnO$_3$
(\emph{R}=Y, Tb, Gd, Dy) \cite{fiebig02, kimura03, goto04} and
TbMn$_2$O$_5$ \cite{hur04}. Among the multiferroic manganates,
\Mn\ takes a special place as it does not contain a rare-earth
ion, leaving Mn formally as the only magnetic ion.\cite{heyer06,
taniguchi06}

Multiferrocity in these materials is explained by the spin current
model,\cite{katsura05a, mostovoy06a} where the electrical
polarization is caused by a spiral order of the magnetic moments.
The spin rotation axis $\mathbf{e}$ does not coincide with the
magnetic propagation vector $\mathbf{Q}$, leading to a
non-vanishing spontaneous polarization $\mathbf{P}\propto
\mathbf{e}\times \mathbf{Q}$. Here we investigate the conditions
for the occurrence of spiral magnetic order responsible for the
rich physics of \Mn. We performed soft x-ray absorption
spectroscopy (XAS) on \Mn\ to determine the local electronic and
magnetic properties of the Mn ions. The results are also compared
to measurements on the isostructural compound \Co, which is a
collinear antiferromagnet and not a multiferroic material. We
analyze the role of the spin-orbit coupling and its impact on the
magnetic structure. The magnitude of the single-ion anisotropy is
calculated using the full-multiplet configuration-interaction
approach. This forms the key to understand why in \Mn\ spiral
magnetic order is possible and why it is suppressed in \Co.

\section{Experimental}

Single crystals of \Mn\ and \Co\ were grown from the melt.
During the single crystal growth process of \Mn\ special care was
taken to keep manganese in divalent state. This was achieved by
the use of high growth temperatures and the renouncement of melt
solvents. Using the top seeded growth technique and low cooling
rate (0.08~K/h), ruby-red transparent single crystals were
obtained starting from a growth temperature of 1574~K.\cite{becker07} 
Single crystals of \Co\ were obtained from
Na$_2$W$_2$O$_7$ melt solution with a small surplus of WO$_3$ (molar ratio CoWO$_4$ : Na$_2$W$_2$O$_7$ : WO$_3$ = 1 : 2 : 0.5),
starting at 1363~K and applying a cooling rate of~3 K/h.

The XAS spectra were collected at the Dragon Beamline of the
National Synchrotron Radiation Research Center in Hsinchu, Taiwan.
The samples were cleaved \emph{in-situ} in an ultra-high vacuum
chamber with pressures in the $10^{-10}$ mbar range, guaranteeing
the high surface quality required to perform bulk-representative
XAS studies in the total electron yield mode. The degree of linear
polarization was $99\pm 1$\%, with an energy resolution of
approximately 0.3~eV. The spectra were collected at room
temperature in the paramagnetic phase of both compounds. Single
crystals of MnO and CoO were measured simultaneously in a separate
chamber as references for \Mn\ and \Co, respectively.

\section{Spectroscopic results}

\begin{figure}[t]
\includegraphics[angle=0,width=9.2cm]{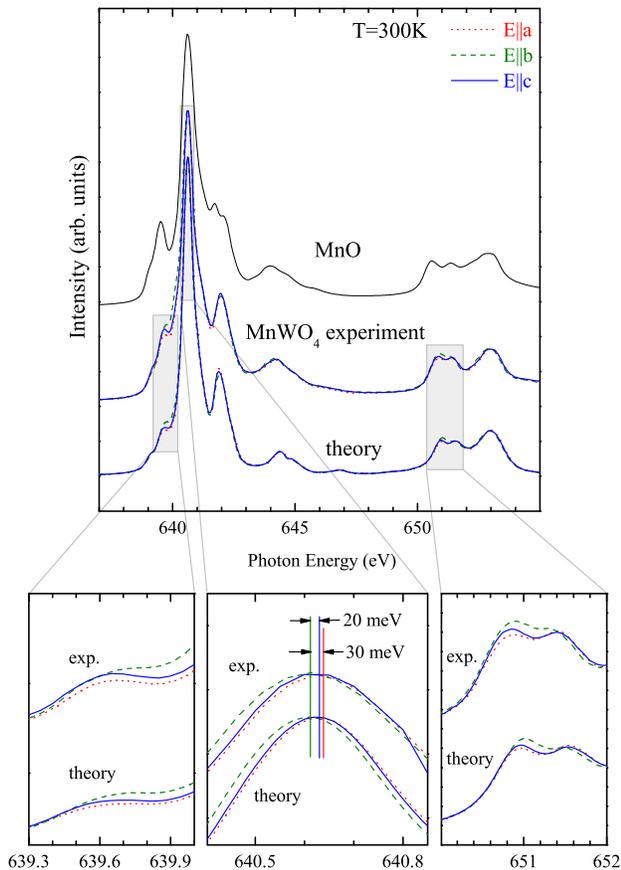}
 \caption[]{(color online)
 Top panel: experimental and theoretical Mn \led\ XAS spectra of \Mn
 with the $\mathbf{E}$ vector of the light parallel to the
 $\mathbf{a}$, $\mathbf{b}$, and $\mathbf{c}$ crystallographic axes.
 The spectrum of MnO is included as reference.
 Bottom panels: a close-up revealing the polarization dependence
 of the spectra.} \label{MnL}
\end{figure}

Fig.~\ref{MnL} shows the room temperature Mn-$L_{2,3}$ XAS spectra
of \Mn\ taken with the  $\mathbf{E}$ vector of the light parallel
to the $\mathbf{a}$, $\mathbf{b}$, and $\mathbf{c}$
crystallographic axes. The spectrum of a MnO single crystal is also
included for reference purposes. The spectra are dominated by the
Mn $2p$ core-hole spin-orbit coupling which splits the spectrum
roughly in two parts, namely the $L_{3}$ ($h\nu \approx$ 638-645~eV) 
and $L_{2}$ ($h\nu \approx$ 649-654~eV) white line regions.
The line shape strongly depends on the multiplet structure given
by the Mn 3$d$-3$d$ and 2$p$-3$d$ Coulomb and exchange
interactions, as well as by the local crystal fields and the
hybridization with the O 2$p$ ligands. Unique to soft XAS is that
the dipole selection rules are very sensitive in determining which
of the 2$p^{5}$3$d^{n+1}$ final states can be reached and with
what intensity, starting from a particular 2$p^{6}$3$d^{n}$
initial state ($n=5$ for Mn$^{2+}$) \cite{degroot94a,tanaka94a}.
This makes the technique extremely sensitive to the symmetry of
the initial state, i.e., the valence and the crystal field states
of the ions.

Comparing the \Mn\ spectra with the MnO, one can immediately
observe that the spectral features are similar with the $L_3$ main peaks at 
identical
energies. This indicates directly that the Mn ions in \Mn\ are also
in the high-spin Mn$^{2+}$ electronic configuration as the Mn ions
in MnO. Yet, the clear differences between the spectra of \Mn\ and
MnO, as seen e.g. at the low and high energy shoulders of the $L_3$ white line,
provide a hint to a crystal field level scheme different from
$O_h$ for the \Mn\ system.

A small but clear polarization dependence for \Mn\ was found, as
can be seen in the bottom panels of Fig.~\ref{MnL}, where a
close-up is shown of the spectra taken with the three different
polarizations. The intensities as well as the energy positions of
several peaks vary with the polarization of the incoming light.
This polarization dependence is yet another indication that the
local symmetry of the Mn ions in \Mn\ is lower than $O_h$.

\begin{figure}[t]
\includegraphics[angle=0,width=8cm]{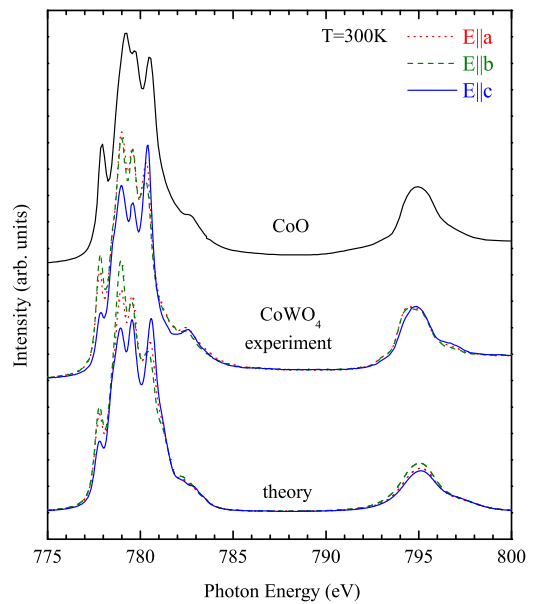}
 \caption[]{(color online) Experimental and theoretical Co \led\ XAS spectra of \Co.
 The spectrum of multi-domain CoO single crystal is included as reference.} \label{CoL}
\end{figure}

Fig.~\ref{CoL} depicts the Co-$L_{2,3}$ XAS spectra of \Co\ taken
with the  $\mathbf{E}$ vector of the light parallel to the
$\mathbf{a}$, $\mathbf{b}$, and $\mathbf{c}$ crystallographic axes,
together with the spectrum of a multi-domain CoO single crystal as
reference. Following the same argumentation as given above for the
Mn \led\ edge, from the strong similarities between the \Co\ and
the CoO spectra we can conclude that \Co\ contains Co$^{2+}$ ions
with a $3d^7$ high-spin configuration like in CoO. In comparing
the \Co\ with the \Mn, one observes that the polarization dependence
is much larger. In the following we will discuss the different
origins of the linear polarization dependence for the two compounds
and its implications for their very different magnetic properties.

\section{Local electronic structure}

To interpret and understand the spectral line shapes and their
polarization dependence, we have performed simulations of the
atomic-like $2p^{6}3d^{n} \rightarrow 2p^{5}3d^{n+1}$ ($n=5$ for
Mn$^{2+}$ and $n=7$ for Co$^{2+}$) transitions using the
well-proven configuration-interaction cluster
model.\cite{tanaka94a,degroot94a,thole97a} Within this method we have
treated the Mn or Co ion within an MnO$_6$ or CoO$_6$ cluster,
respectively, which includes the full atomic multiplet theory and
the local effects of the solid. It accounts for the intra-atomic
$3d$--$3d$ and $2p$--$3d$ Coulomb interactions, the atomic $2p$
and $3d$ spin-orbit couplings, the local crystal field, and the
O~$2p$--Mn~$3d$ or O~$2p$--Co~$3d$ hybridization. This
hybridization is taken into account by adding the
$3d^{n+1}\underline{L}$ and $3d^{n+2}\underline{L}^{2}$ etc.
states to the starting $3d^{n}$ configuration, where
$\underline{L}$ denotes a hole in the O $p$ ligands.

\begin{figure}[t]
\includegraphics[angle=0,width=8cm]{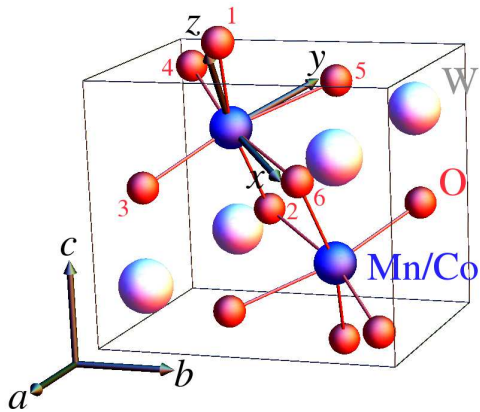}
  \caption[]{(color online) The structure of \Mn\ and \Co.
  For the calculations of the XAS spectra, a local coordinate system
  ($x$, $y$, $z$) was introduced, in which the axes point roughly
  along the Mn/Co-O bonds. The global crystallographic coordinates
  system can be transformed into the local coordinate system by rotating
  along $b$ by $\approx 34^\circ$ and then along the $c$ axis by $45^\circ$
  (see text for more details).} \label{structure}
\end{figure}

To facilitate the setup and the interpretation of the
calculations, a local coordinate system was introduced, with its
axes ($x$, $y$, $z$) pointing roughly along the Mn-O and Co-O
bonds. To transform the global coordinates ($a$, $b$, $c$) into
the local system, first a rotation by 33.9$^\circ$ along $b$ is
applied, resulting in a new system with ($a^\prime$, $b^\prime$,
$c^\prime$). This system is then rotated by 45$^\circ$ along the
\emph{old} axis $c$, transforming ($a^\prime$, $b^\prime$,
$c^\prime$) into ($x$, $y$, $z$). The local coordinate system is
shown in Fig.~\ref{structure}. The coordinates of the oxygen ions
as well as the Mn-O and Co-O distances are summarized in
Table~\ref{ltab}. The crystal structure has been taken from
Ref.~\onlinecite{lautenschlaeger93}. There are two distorted octahedra in
the unit cell and they are connected to each other by inversion,
thereby contributing to the polarization dependence in the same
manner.

\begin{table}
\begin{tabular}{ccccccc}
\hline
MnWO$_4$ \\
\hline O & \hspace{3mm}x\hspace{3mm} & \hspace{3mm}y\hspace{3mm} &
\hspace{3mm}z\hspace{3mm} & Mn-O dist. &
\hspace{3mm}$pd\sigma$\hspace{3mm} & \hspace{3mm}
$pd\pi$\hspace{3mm} \\
1 & 0.25 & 0.24 & 2.15 & 2.18 & -1.18 & 0.54 \\
2 & 0.25 & 0.24 & -2.15 & 2.18 & -1.18 & 0.54 \\
3 & 0.31 & -2.07 & -0.02 & 2.09 & -1.36 & 0.63 \\
4 & -2.06 & 0.36 & 0.02 & 2.09 & -1.36 & 0.63 \\
5 & 0.31 & 2.24 & -0.08 & 2.27 & -1.02 & 0.47 \\
6 & 2.25 & 0.25 & 0.08 & 2.27 &  -1.02 & 0.47 \\
 &  \\
\hline
CoWO$_4$ \\
\hline
O & x & y & z & Co-O dist. & $pd\sigma$ & $pd\pi$ \\
1 & 0.19 & 0.19 &  2.10 & 2.12 & -1.07 & 0.49 \\
2 & 0.19 & 0.19 & -2.10 & 2.12 & -1.07 & 0.49 \\
3 & 0.24 & -2.02 & -0.04 & 2.03 & -1.24 & 0.57 \\
4 & -2.02 & 0.25 & 0.04 & 2.03 & -1.24 & 0.57 \\
5 & 0.24 &  2.14 & -0.03 & 2.16 & -1.01 & 0.47 \\
6 & 2.14 & 0.23 & 0.03 & 2.16 & -1.01 & 0.47 \\
\end{tabular}
\caption[]{Coordination of the \Mnz\ and \Coz\ ions in their
distorted oxygen octahedra. All oxygen positions are given in \AA\ in the
local coordinate system specified in fig.~\ref{structure}, where
the numbering of the atoms is also defined. The hybridization
strengths $pd\sigma$ and $pd\pi$ are in units of eV.} \label{ltab}
\end{table}

The Mn-O and Co-O bond lengths were used to estimate the
hybridization strength using Harrison's
description,\cite{Harrison} resulting in values for $pd\sigma$ and
$pd\pi$ as shown in Table~\ref{ltab}. The Slater-Koster
formalism\cite{slater54} provides the angular dependence of the
hybridization strength. Values for the crystal fields were tuned
to find the best match to the experimental spectra. Parameters for
the multipole part of the Coulomb interactions were given by the
Hartree-Fock values,\cite{tanaka94a} while the monopole parts
($U_{dd}$, $U_{pd}$) were estimated from photoemission experiments
on MnO\cite{bocquet92a} and previous work on CoO.\cite{csiszar05a}
The simulations were carried out using the program XTLS~8.3,\cite{tanaka94a}
and the parameters used are listed in Refs.~\onlinecite{Mn2para, Co2para}.

The calculated polarization-dependent spectra for the Mn \led\
edge of \Mn\ are plotted in Fig.~\ref{MnL}. One can observe that
the general line shape of the experimental spectra is very well
reproduced. Equally important, also the small energy shifts in the
peak positions as well as the small variations in the peak
intensities as a function of polarization can all be simulated.
This indicates that we have been able to capture the local
electronic structure of the Mn ion in \Mn\ with great accuracy. We
now will look into this in more detail.

\begin{figure}[t]
\includegraphics[angle=0,width=9cm]{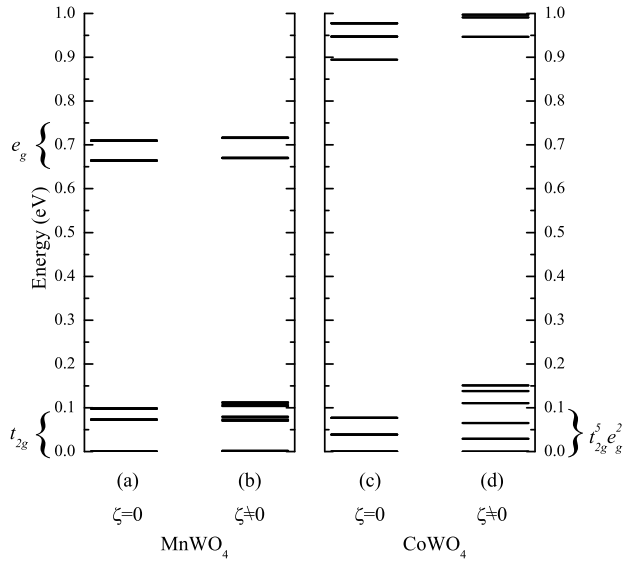}
\caption[]{Energy level diagrams for (a,b) Mn$^+$ ($3d^6$) and
(c,d) \Coz\ clusters, excluding and including the $3d$ spin-orbit
coupling $\zeta$. Only the states up to 1~eV are shown.}
\label{elevel}
\end{figure}

The ground state of a Mn$^{2+}$ ion in $O_h$ symmetry is $^6A_1$,
and this will not split or change upon lowering the crystal
symmetry. The electronic charge distribution of this high-spin
half-filled shell ion remains spherical. The presence of lower
symmetry crystal fields will then show up most clearly when an
electron is added (or removed) from the valence shell. To
illustrate such a situation, we depict in the left panel of
Fig.~\ref{elevel} the total energy level diagram of a Mn$^+$
($3d^6$) ion using the crystal and ligand field parameters which
reproduce best the Mn \led\ spectra of \Mn\ as described above.
One can readily observe a splitting of about 0.7 eV between states
with the added electron in the \tg\ vs. the \eg\ orbitals. Upon
switching off the $3d$ spin-orbit interaction, see
Fig.~\ref{elevel} (a), one can also identify an intra-\tg\
splitting of about 65 and 95~meV as well as an intra-\eg\
splitting of 50 meV. All of the orbitals are of mixed type as a
result of the low local symmetry. The two orbitals lowest in
energy are approximately $1/\sqrt{2}(xz+yz)$ and $1/\sqrt{2}(xz-yz)$, resembling
the $xz$ and $yz$ orbitals, rotated by $\pm$45$^\circ$ along $z$.
This is a result of the local $S_2$ symmetry axis of the MnO$_6$
cluster, which lies in the $xy$ plane with an angle of roughly
45$^\circ$ to the oxygen bonds.

Also in the Mn \led\ XAS process one extra electron is added into
the valence shell. The lowest lying peak at 639.6~eV, see Fig.~\ref{MnL}, 
involves the excitation of the $2p$ core
electron into the \tg\ orbitals. Different polarizations access
different orbitals with different probabilities. For the 640.6 eV
peak, the polarization dependence is reflected in terms of shifts
in its energy position. If one could have measured the
polarization dependence along the local coordinates of one Mn ion,
one may expect to observe peak shifts of 65 and 95~meV associated
with the intra~\tg\ splittings shown in Fig.~\ref{elevel}. Yet,
having two Mn ions in the unit cell, and measuring along the
global $a$, $b$, and $c$ crystallographic directions, the shifts
between the effectively composite peaks become reduced. In the
experiments and in the simulations we are left with shifts of
about 20 and 30 meV as can be seen in the middle bottom panel of
Fig.~\ref{MnL}. For the 639.6 eV peak, the shifts due to variation
of the polarization are more difficult to quantify since there is
also a variation in the intensity of the peaks. The latter is
apparently caused by the rather complicated multiplet effects
involving the $2p$ core hole, as demonstrated by the excellence of
the match by the simulations, see the left bottom panel of
Fig.~\ref{MnL}. These intensity variations are yet quite small,
and \textit{integrated} over the entire \led\ range, would have
been even identical to zero if the non-cubic hybridization effects
were absent.

The simulations for the polarization dependent spectra for the Co
\led\ edge of \Co\ are shown in Fig.~\ref{CoL}. Like in the Mn
case, we have been able to achieve a satisfying fit to the
experimental spectra: all features are well reproduced,
including the strong polarization dependence. We will now analyze
the local electronic structure of the Co ion on the basis of the
parameters used in these simulations.

For the Co$^{2+}$ ions, the ground state of the $3d^7$
configuration in $O_h$ symmetry is $^4T_1$ ($\approx
t_{2g}^5e_g^2$). Unlike for the $^6A_1$ of the Mn, this $^4T_1$
state will be split upon going to lower crystal symmetry. The
degeneracy in the orbital part, i.e. in the $t_{2g}$ sub-shell,
will be lifted. This can be seen from the total energy level
diagram shown in the right panel of Fig.~\ref{elevel}. The
calculation with the $3d$ spin-orbit interaction switched off, see
Fig.~\ref{elevel} (c), reveals the presence of three low-lying
quartets within the first 0.1 eV. In an one-electron language,
each of them would correspond to the hole occupying one of the
three \tg orbitals. The intra-\tg\ splittings are 40 meV and 80
meV. These values are quite close to those of the \Mn\ case
reflecting the similar crystal structure. The importance of the
lifted degeneracy is that the orbital occupation will no longer be
isotropic, i.e. the Co ion charge distribution will be highly
non-spherical. This then explains the strong polarization
dependence in the absorption spectra.

It is also important to notice how the low-lying states of \Coz\
are influenced by the $3d$ spin-orbit coupling. The energy level
diagram in Fig.~\ref{elevel} (d) reveals that there is indeed a
large amount of mixing due to spin-orbit interaction. The
spin-orbit coupling constant of $\zeta=66$ meV is of the same
order of magnitude as the splittings in the \tg\ subshell. All
this has direct consequences on the magnetism, as will be
discussed in the next section.

\begin{figure}
 \includegraphics[angle=0,width=6cm]{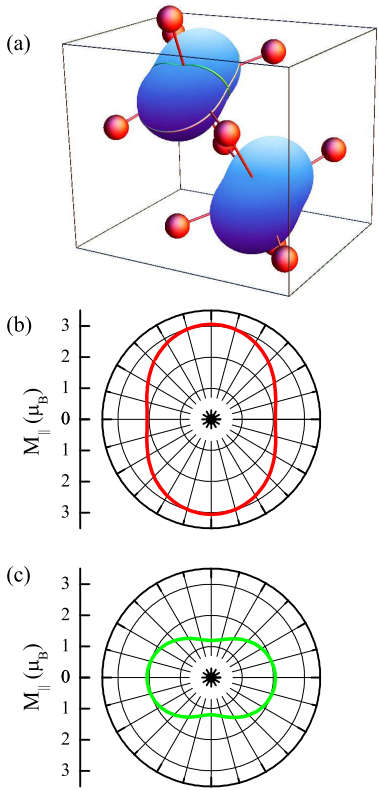}
 \caption[]{(color online) Calculated magnetic anisotropy of \Coz\ in \Co.
(a) Three-dimensional representation of the magnetic anisotropy in
the unit cell. The surface depicts the expectation value of the
magnetic moment $M_\parallel$ pointing parallel to a small
exchange field applied to the ion. (b) Two-dimensional cut of
$M_\parallel$ in the plane of easy and intermediate magnetic axis
(corresponding approximately to the $xy$ plane in local
coordinates). (c) Cut through the plane of hard and intermediate
magnetic axis.} 
\label{dir}
\end{figure}

\section{Single-Ion anisotropy}

From the results of the simulation, we find that the $3d$
spin-orbit interaction has a major influence on the energies and
nature of the low lying states of the \Co\ system. The magnetic moment will then not
only have a spin contribution but also an appreciable orbital one.
This results in a strong coupling of the direction of magnetic moment to the crystal structure.
We can make an estimate of the single ion anisotropy of the Co$^{2+}$
ion. For that we use the same parameters with which we
have reproduced the XAS spectra and the polarization dependence
therein, and we apply a small exchange field of $\mu_B
H_{ex}=10^{-6}$ eV ($H_{ex}\approx 200$ Oe) to align the magnetic
moment. Here we would like to note that the results hardly changes
with increasing the strength of the field, i.e. they are robust
even beyond $\mu_B H_{ex}=10^{-2}$ eV.

Fig.~\ref{dir} shows the expectation value of the magnetization
parallel to the applied exchange field $M_\parallel$ as a
three-dimensional representation and in a polar plot, the latter
in local coordinates. It can be seen that the single-ion
anisotropy is of easy-axis type, with the axis lying along $\phi=$
135$^\circ$ at $\theta=90^\circ$ in the local coordinate system.
In the direction of the easy axis, the moment is completely
directed along the magnetic field, and the perpendicular component
of the magnetization vanishes. The easy axis is also perpendicular
to the $S_2$ symmetry axis of the distorted octahedron, reflecting the
orbital character of the two lower lying ligand field levels. The
total magnetic moments of the hard and easy axes are 1.2~$\mu_B$
and 3.0~$\mu_B$, respectively, with an orbital contribution of 0.1~$\mu_B$
and 0.9~$\mu_B$.
The easy axis of the total moment
coincides directly with that of the orbital moment, which is also
pointed out earlier by Bruno.\cite{bruno89} It has also been
reported that for low symmetries, the orbital and total moment
could differ in their behavior due to the magnetic dipole moment
$T_z$.\cite{vanderlaan98} In the case of \Co, however, with the
symmetry of the CoO$_6$ groups being not too far from cubic, this $T_z$ is around
$-0.01$ eV, and does not influence the anisotropy significantly.

In \Co, the single-ion anisotropy is strong enough to lead to an
effective pinning of the spins in the magnetically ordered phase.
Isotropic superexchange will be the dominating magnetic
interaction, although in addition anisotropic exchange is
possible, as the bonds lack inversion centers. But the
Dzyaloshinskii-Moriya (DM) term
$\mathbf{D}\cdot(\mathbf{S_1}\times\mathbf{S_2})$, that favors
canted magnetic moments, is small compared to the energy involved
in the magnetic anisotropy of \Co, reflected by the large difference
in moment of more than 1~$\mu_B$. The superexchange will then
minimize the energy in the magnetically ordered phase by aligning
the moments along the easy axis. The easy axis is identical for
both CoO$_6$ clusters in the unit cell, and thus the single-ion
anisotropy itself will not lead to canting between the magnetic
moments of the \Coz\ ions. The result is a collinear
antiferromagnet with an easy axis approximately in the $ac$ plane
$\approx 40^\circ$ off from $a$. This direction is illustrated in
Fig.~\ref{dir} and is close to the one found in the experiment by
neutron diffraction. \cite{weitzel77a}

Returning to \Mn\ case: to the leading order, the local spherical
$^6A_1$ ground state does not produce an orbital moment. The spin
is then free to point in any direction. This is reproduced in the
calculation, where the response of the magnetic moment to an
exchange field is practically isotropic. Yet, in second order, the
spin-orbit coupling does affect the Mn-O hopping. This produces a
non-vanishing moment perpendicular to the exchange field, albeit
in the order of 10$^{-2}$~$\mu_B$, being much lower than in the
case of the spin-orbit active \Co. Consequently, the DM term is on
the same energy scale as the anisotropy, and spin canting can
easily occur. The propagation direction of the spiral magnetic
order is in the \emph{local} $xy$ plane,
meaning that the second order process leads to a hard axis along
$z$. This demonstrates the importance of the single-ion anisotropy
in this system. The form of the anisotropy, however, depends
strongly on the magnetic exchange. This explains that even three
different magnetically ordered phases exist in the material at low
temperatures.\cite{ehrenberg97} Additionally to the hard axis, the easy plane also
has two principal axes for the magnetic anisotropy with a small
difference between the two. This, in combination with the subtle
dependency of the anisotropy on the exchange field, makes
incommensurabilities likely to appear.

\section{Conclusion}
The single-ion anisotropies extracted from our soft x-ray
absorption data and the corresponding full-multiplet calculations
provide a natural explanation for the magnetic ordering phenomena
occurring in \Co\ and \Mn. In \Co, the spin-orbit coupling and a
non-vanishing orbital moment overwhelm the anisotropic Dzyaloshinskii-Moriya exchange
and cause the formation of a collinear ordering of the moments.
Contrary to that, although being isostructural, \Mn\ shows only a
small magnetic anisotropy which is comparable in energy to the Dzyaloshinskii-Moriya
interactions. This will cant the spins in such a way that the
spiral propagates perpendicular to the direction of smallest
magnetic moment, allowing a modulated spin spiral which causes the
ferroelectricity in the material.

\section{Acknowledgements}
We gratefully acknowledge the NSRRC staff for providing us with
beamtime. The research in Cologne is supported by the Deutsche
Forschungsgemeinschaft through SFB 608. N. H. is also supported by
the Bonn-Cologne Graduate School of Physics and Astronomy. We are
grateful for discussions with M.~W.~Haverkort and D.~I.~Khomskii.

\end{document}